**RecA and RecB: probing complexes of DNA repair proteins with mitomycin C in live Escherichia coli with single-molecule sensitivity**

Alex L. Payne-Dwyer[1,2,†], Aisha H. Syeda[1,2,†], Jack W. Shepherd[1,2], Lewis Frame[3] and Mark. C. Leake[1,2*]

[1] Department of Physics, University of York, York, UK, YO10 5DD;

[2] Department of Biology, University of York, York, UK, YO10 5DD;

[3] School of Natural Sciences, University of York, York, UK, YO10 5DD.

[*]Correspondence: mark.leake@york.ac.uk

[†]These authors contributed equally.

**Abstract**

The RecA protein and RecBCD complex are key bacterial components for the maintenance and repair of DNA, RecBCD a helicase-nuclease that uses homologous recombination to resolve double-stranded DNA breaks and also facilitating decoration of single-stranded DNA with RecA to form RecA filaments, a vital step in the double-stranded break DNA repair pathway. However, questions remain about the mechanistic roles of RecA and RecBCD in live cells. Here, we use millisecond super-resolved fluorescence microscopy to pinpoint the spatial localization of fluorescent reporters of RecA and the RecB at physiological levels of expression in individual live *Escherichia coli* cells. By introducing the DNA crosslinker mitomycin C, we induce DNA damage and quantify the resulting changes in stoichiometry, copy number and molecular mobilities of RecA and RecB. We find that both proteins accumulate in molecular hotspots to effect repair, resulting in RecA filamental stoichiometries equivalent to several hundred molecules that act largely in RecA tetramers before DNA damage, but switch to approximately hexameric subunits when mature filaments are formed. Unexpectedly, we find that the physiologically predominant form of RecB is a dimer.

**1. Introduction**

Accurate duplication of the genome is crucial in all organisms, accomplished by a sophisticated molecular machine of the replisome (1). An inability to accurately replicate genetic material can lead to cell death and/or cancers (2, 3). Mitomycin C (MMC) is a naturally occurring antibiotic that can be used to controllably disrupt DNA replication, and thus a valuable reagent in studying DNA repair processes. It finds major use as a chemotherapeutic in treating several cancers (4) and retinopathies (5). It acts by targeting DNA deoxyguanosine (dG) residues (6), forming intrastrand or interstrand crosslinks (7). If unrepaired, these can interfere with cellular processes such as transcription and replication, potentially causing genome instability (8). An encounter between a mitomycin C-induced crosslink and an approaching replisome, may lead to replisome disassembly and eventually to a double strand break (DSB) (9). DSBs are recognised by RecBCD in *E. coli* (10). The ends are then processed by RecBCD to generate 3'-ended single stranded DNA (ssDNA), potentially creating a landing pad for the principal recombination protein, RecA (10). Recombination of RecA-ssDNA with the homologous DNA restores the replication fork, on which the replisome can be reloaded. The replisome may resume replication if the blocking adduct is repaired (11).

RecBCD forms parts of two major pathways for homologous DNA recombination, essential for DSB repair (10). RecBCD activities involve several processes - it recognises and binds DSBs, begins unwinding both DNA strands, and also degrades both (10). This activity continues unhindered until it encounters an octameric Chi site that induces a shift in enzyme activity to degrade only 5'-ended strand (12, 13). This activity shift results in a 3'-ended ssDNA overhang that facilitates RecA loading. RecA-ssDNA nucleoprotein filaments can infiltrate an intact duplex and, on finding homology, recombine with the infiltrated duplex (14, 15). Further processing of the resulting structure and the action of replication-reloading primosome proteins help in the establishment of an intact replisome to resume replication (16). Recombination proteins, such as RecBCD, need access to replication-transcription conflict sites and collapsed forks, but if RecBCD is missing then dsDNA is degraded by exonucleases (17, 18), possibly resulting from replisome disassembly. How RecA stabilizes blocked forks is still an open question.

Here, we use millisecond super-resolved Slimfield microscopy (19) in live *E. coli* of genomically-encoded fluorescent fusions RecA-mGFP (20) and RecB-sfGFP (21) expressed from their native promoters. Since RecA fusion constructs retain only partial function, our approach makes use of a merodiploid RecA fusion that express from one copy of the native gene and one copy of the fusion construct under the same promoter (20). Slimfield uses sampling timescales that are sufficiently rapid to enable direct tracking of diffusing fluorescently labeled proteins in the cytoplasm of living cells (22). This technology enables an internal calibration for molecular counting by quantifying the brightness of a single fluorescent protein *in vivo*, and additionally offers trajectory information at typically 40 nm lateral precision. We analyse Slimfield images to reveal the density and spatial distribution of RecA and RecB in individual cells. We identify assemblies of RecA, and find that the number of molecules associated with assemblies has a characteristic 'repeat unit' that increases between cellular states of SOS readiness and MMC-induced response, while the equivalent repeat unit of RecB assemblies is unaffected. We also measured changes in mobility and localization of RecA and RecB upon treatment with a sublethal MMC dose just short of inducing cell filamentation. These changes include formation of RecA filaments associated with SOS and the aggregates that precede it. Imaging RecA presents far brighter fluorescent foci than RecB indicating a cellular concentration 2-3 orders of magnitude greater. We observed an increase in RecA copy number not RecB upon MMC treatment, with up to 20% of cells devoid of RecB assemblies. Our results shed new light on links between structure and function for RecA and RecBCD in mediating repair upon DNA damage.

## 2. Results

*2.1 Abundance of RecA, but not RecB, increases on MMC-induced DNA damage*

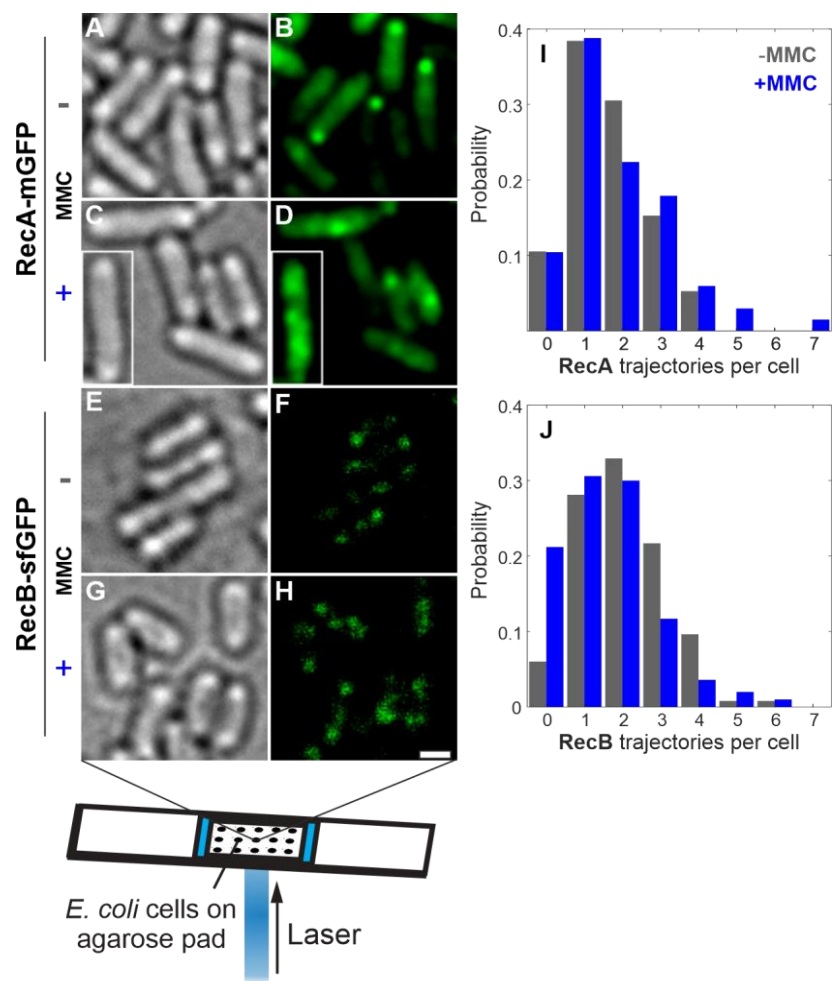

**Figure 1.** Brightfield and Slimfield of live *E. coli* in minimal media, labelled at RecA-mGFP or RecB-sfGFP before and after MMC treatment. Inset (C,D) is zoom-in from same field of view. Brightness of RecB-GFP Slimfield panels (F,H) scaled 100x *vs.* RecA-GFP panels (B,D). Scale bar 1 μm. (I,J) Probability distributions for number of tracks detected per cell.

We find that in the absence of MMC RecA-GFP has an an approximately uniform distribution in the cytoplasm occasionally punctuated with bright fluorescent foci (Figure 1B). By integrating pixel fluorescence intensities in each cell (23) and normalising by the brightness of a single GFP (24) we find the total amount of RecA-mGFP increases from 11,500 ± 200 molecules (±SEM) in untreated cells to 14,100 ± 300 molecules in MMC treated cells (Figure S2a). MMC treatment caused the subset of RecA-mGFP molecules localised in tracks to double from 510 ± 30 to 1,080 ± 60 per cell (Table S1).

RecB-GFP also exhibited fluorescent foci on a diffuse background, before and during MMC treatment (Figure 1F,H). The RecB copy number is sufficiently small that estimates based on total cellular fluorescence (23) must be quantitatively distinguished from total autofluorescence, which we estimate from unlabelled MG1655 parental cells grown and imaged under identical conditions. We find the mean level of RecB-associated fluorescence, including the captured RecB tracks, is almost three times greater than the cell autofluorescence. Thus, we cannot consider all of the cellular RecB as residing in trajectories and we conclude that a significant cytoplasmic pool of RecB exists that is sufficiently mobile to evade direct particle-tracking-based detection, in particular using slower sampled images from commerical microscope systems.

The RecB copy number reduces following MMC-induced DNA damage, comprising 126 ± 11 molecules per cell before treatment and 101 ± 14 molecules afterward (Figure S2B, significant by Brunner-Munzel (BM) test, n=246, p=0.0216), without a measurable change in cell size. Of these, the mean number of RecB localised in tracks is relatively small but more sensitive to MMC than RecA; just 13.6 ± 0.5 molecules per cell in all tracks, decreasing to 9.3 ± 0.3 on treatment.

*2.2 RecB forms characteristic puncta which are partially lost on MMC exposure*

1-3 trajectories of RecA or RecB were detected in each cell above the local background (Figure 1I,J) . However, RecA and RecB showed strongly opposing trends in the number of tracks observed upon MMC treatment. While RecA showed no significant change from 1.66 ± 0.06 to 1.86 ± 0.16 trajectories detected per cell on MMC treatment (BM test, n=60, p=0.50), the population-average number of RecB tracks was significantly reduced from 2.06 ± 0.09 to 1.56 ± 0.06 per cell. If, however, we set aside the fraction of cells with no detected RecB tracks, the mean number of RecB tracks falls, from 2.19 ± 0.10 to 1.98 ± 0.07 tracks (BM test, n=234, p<0.001) containing 12.1 ± 0.3 molecules per cell. This unexpected population devoid of RecB foci otherwise resembles the other treated cells; rather than filamenting, they are 8 ± 3% shorter on average and retain the same cytoplasmic pool of RecB.

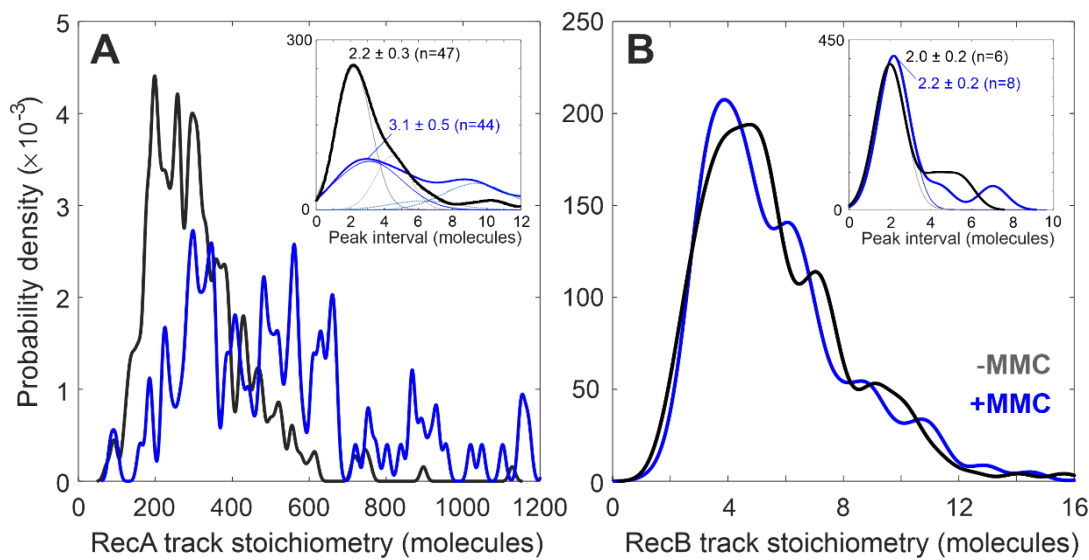

**Figure 2.** Stoichiometry of detected foci of RecA-mGFP and RecB-sfGFP with (blue) or without mitomycin C treatment (black), shown as kernel density estimates *(25)* with width of 8 RecA molecules for clarity, or kernel width of 0.7 molecules, following the known detection sensitivity to single GFP. Insets are the distributions of intervals between nearest neighbour stoichiometry peaks identified at a kernel width of 0.7 molecules (solid curves) that is a signature for fundamental subunit stoichiometry. Overlaid are the constrained Gaussian fits that minimize a reduced χ2 metric, with components of equal width and whose centres are fixed at integer multiples. The resulting fits comprise three components for RecA with MMC treatment (blue, Pearson's $R^2$ = 0.979, 5 degrees of freedom (dof)) and two components for RecA without MMC (grey, Pearson's $R^2$ = 0.961, 4 dof). The mode of the peak interval is indicated ± 95% confidence interval, with the number of contributing peak pairs in the original stoichiometry distribution.

Foci in RecA were two orders of magnitude more intense than those of RecB, corresponding to a greater apparent stoichiometry. On MMC treatment, the mean RecA-mGFP stoichiometry almost doubled from 310 ± 8 to 580 ± 30 molecules per focus, reflecting local accumulation of RecA protein (Figure 2A). As the RecA-mGFP strain is merodiploid, the *recA-mgfp* gene fusion construct expressed from the same promoter as the unlabelled endogenous *recA* gene (20), the total amount of cellular RecA is approximately twice the detected amount based on fluorescence. This tendency for localised accumulation of RecA hotspots is consistent with prior observations of long nucleoprotein filament formation on single stranded DNA (20). Conversely, the extended filaments we see upon MMC treatment may indicate increased occurrence of processed ssDNA. We find tyypically ~30 RecA in the 'pool' may comprise untracked diffusive molecules (Figure S2C).

The mean RecB stoichiometry falls from 6.6 ± 0.1 to 6.1 ± 0.2 molecules per focus (BM, n=478, $p<10^{-6}$) (Figure 2B). A mean of ~6 RecB is reconciled by comprising 3 copies of a subunit whose stoichiometry is 2 molecules (Figure 2B inset) . RecB foci are observed more commonly near the poles of the cell (Figure 1G-H). A pool stoichiometry of ~1 molecule (Figure S2B,D) suggests that the untracked majority of all RecB (90% ± 1%) are monomers irrespective of MMC treatment (BM test, n=243, p=0.27).

*2.3. RecA reorganises into filaments with approximately hexameric subunits in response to MMC*

RecA and RecB stoichiometry distributions show clear gaps (Figure 2A, 2B & insets). One explanation is that each detected fluorescent focus has a diffracted-limited width of ~250 nm that may potentially contain more than one subunit of RecA or RecB, such that the measured focus stoichiometry may appear as an integer multiple of the subunit stoichiometry, manifest as periodic peaks on the focus stoichiometry distribution. The difference between pairs of values on the stoichiometry distribution is, within measurement error, either zero or an integer multiple of the subunit stoichiometry. The magnitude of the most likely non-zero pairwise difference value corresponds to the subunit stoichiometry, with less likely values corresponding to harmonic peaks. RecA has a most likely pairwise interval value of 2.2 ± 0.3 RecA-mGFP molecules before MMC addition; accounting for merodiploid half-labelling, this indicates 4.4 ± 0.5 RecA molecules in total per subunit. After MMC treatment, the most likely interval value is 3.1 ± 0.5 RecA-mGFP molecules, implying a subunit comprising 6.2 ± 1.0 RecA molecules in total.

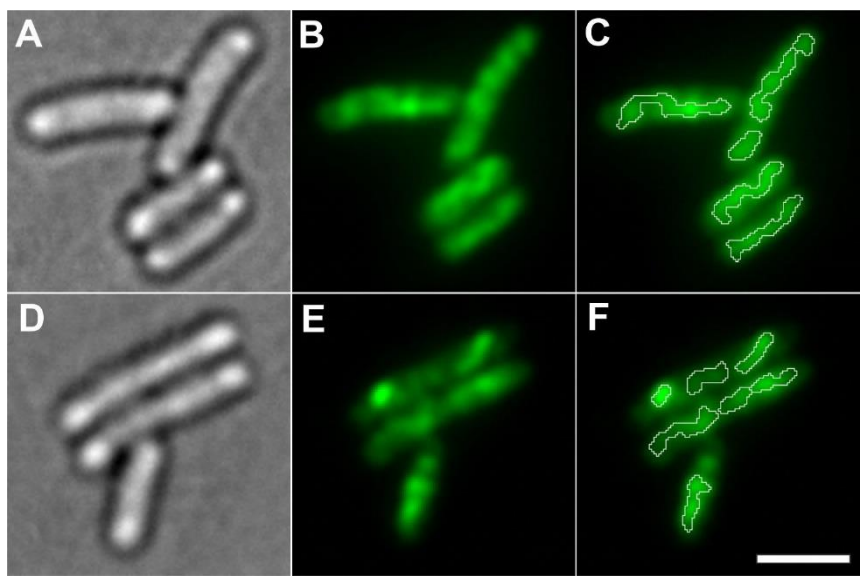

**Figure 3.** Filamentous bundles of RecA-mGFP in MMC treated cells as observed in A,D) brightfield and B,E) Slimfield, shown with C,F) indicative local segmentation of the bundles (white overlay). Scale bar 2 μm.

In MMC-treated cultures we observe strikingly bright, bundle-like structures (Figures 1D and 3B-E), >95% of which exceed a local concentration of twice the mean stoichiometry in untreated cells (Figure S3).

The mean number of segments containing MMC-induced bundles is 1.9 ± 0.3 per cell, which is consistent with a pair of RecA bundles extending along each cell, possibly indicating a single double stranded DNA (dsDNA) break with sister homology consistent with recent findings relating to the SMC protein RecN influencing RecA translocation and remodelling *(26)*. The copy number of filamented RecA-mGFP is 2,600 ± 190 molecules per MMC-induced bundle, but each occurs in consistent multiples of approximately 1,500 ± 300 molecules (Figure S3). When ATP is present, the binding site density on each helical filament containing ssDNA was found in previous studies to be 1.5 nm per RecA *(27, 28)*, here of which approximately half is unlabelled. The filaments are known to be undersaturated with RecA under physiological conditions *(29)*. We find that each MMC-induced bundle typically fills a region measuring 2.0 ± 0.9 µm (mean ± s.d.) long, up to 0.7 ± 0.2 µm wide and ~0.6 µm deep, but contains a qunatity of RecA equivalent to >10 µm total length of helical filament.

In contrast, the brightest RecA assemblies in untreated cells occur in isolation, and are never elongated but reside within diffraction-limited foci (Figure 1B) which may be related to RecA storage bodies observed previously *(30)*. Assemblies whose RecA content exceeds twice the untreated stoichiometry (10 ± 3% of untreated cells) and typically contain 800 ± 500 RecA (Figure S3). This quantity is equivalent to >2 µm of filament inside sphere <0.4 µm in diameter. Our results are in agreement with a recent observation of RecA filaments in *Caulobacter crescentus* cells (diameter 1.0 ± µm, 5-6 µm long) *(26)*.

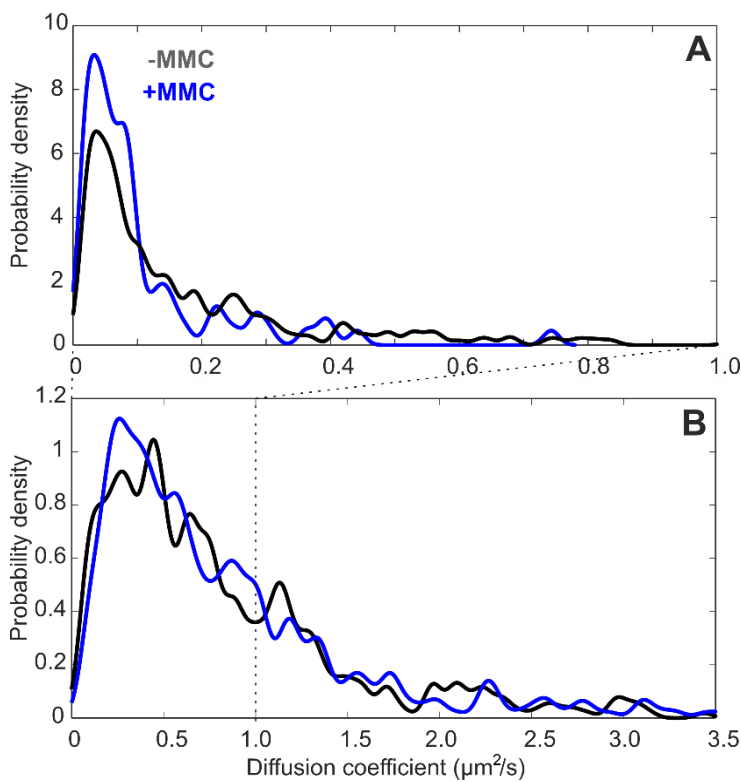

**Figure 4.** Kernel density distributions of instantaneous microscopic diffusion coefficient for tracks of A) RecA-mGFP and B) RecB-sfGFP. The mean diffusion coefficient of tracked RecA decreases from 0.17 ± 0.02 µm$^2$/s to 0.07 ± 0.01 µm$^2$/s on MMC treatment (Figure 4A). This change in diffusivity likely reflects the proportion of RecA condensed onto ssDNA as filaments. In contrast, the mean diffusion coefficient of tracked RecB is unaffected by MMC (Figure 4B), with untreated and treated values of 0.82 ± 0.03 µm$^2$/s and 0.79 ± 0.03 µm$^2$/s respectively (BM test, n=478, p=0.48).

## 3. Discussion

Here, we investigated the stoichiometry and distribution of fluorescently tagged RecA and RecB proteins in live *E. coli* upon treatment with the DNA alkylating agent MMC. We developed an image analysis pipeline incorporating image segmentation driven by machine-learning (Figure S5). We show that our sensitivity and dynamic range are sufficient to quantify counts, either by stepwise photobleaching of multi-molecular complexes or by direct detection of single molecules at sufficient speed. Our findings show that the RecA copy number increases upon treatment with MMC. We observed a modest increase in the number of RecA tracks concomitant with almost doubling of stoichiometry per focus when compared with untreated cultures. This tendency for localized concentration of RecA indicates significant assembly formation and is consistent with the known property of RecA to form long nucleoprotein filaments on ssDNA as nucleated from polar locations (20). We observed extended filamentous bundles in MMC treated cultures, possibly due to increased availability of processed ssDNA. Bundles were preferentially aligned along the cell axis, which cannot be explained only as a result of elongation under confinement inside the cell; the observation of more extended filaments on MMC treatment may suggest search processing for more distant sites of homology, a lower total success rate of homology search due to a higher incidence of DNA breaks, or obstructive interactions between RecA bundles corresponding to simultaneous different DNA breaks.

RecA assembly formation is not limited to conditions of excess DNA damage. Before treatment, a portion of RecA appears in dense polar foci. These occasionally bright foci may indicate stochastic DNA damage in a proportion of cells of an otherwise healthy culture, but can be compared with the storage structures described previously, that are functionally distinct from filaments (30). However, there is strong evidence for bundle nucleation at the cell membrane (31). Our results, including observation of an average of ~2 RecA clusters per cell, are consistent with RecA foci at poles providing a nucleation point for filament formation.

We observed a change in the subunit stoichiometry for RecA foci from ~4-mer in untreated cells, to a ~6-mer within spatially extended filaments following MMC treatment. Previous reports indicate that RecA undergoes linear polymerization in a head-to-tail fashion with stable trimeric, tetrameric and hexameric forms when ssDNA is present (32). In vitro and in vivo studies provide strong evidence for dimeric nucleation points on ssDNA mediated by SSB (33) leading to bidirectional, diffusion-limited growth (34). Our snapshot observation of filament stoichiometry cannot shed light directly on models of dynamic nucleation or stepwise growth. Rather, it indicates the relative stability of characteristic subunits within the mature filament. Each subunit has a split-ring structure related to the intact hexameric ring of DNA helicases, but distorted axially such that rings each complete a single helical turn around ssDNA (35). One hypothesis is that this RecA patterning could result from a small, periodic barrier to further polymerization. While our stoichiometry analysis suggests variability across different foci, our results favour a ~6-mer subunit, consistent with the hexameric form of RecA in vitro. Although these data cannot directly establish whether independent hexamers of RecA occur in vivo either on DNA or in the cytosol, it is conceivable that assembly and rearrangement of RecA hexameric subunits on DNA could generate the canonical ATP-inactive and ATP-active DNA-binding filaments (36, 37). In light of a recent study highlighting the role of RecN in RecA filament formation and activity (26), we hypothesize that the tetrameric form of RecA may be devoid of RecN and is ATP-inactive, which changes to an approximately hexameric form upon DNA damage via the involvement of RecN and its associated ATP-activity.

RecA bundles are drawn from a cytoplasmic pool on a timescale of 5-15 min after UV irradiation (20), with factors such as RecBCD, typically enabling homologous recombination within 2h (20). The large increase we detect in copy number and decrease in diffusivity, but with similar pool concentration before and 3h after addition of MMC, indicate that the RecA pool reservoir is restored to the same steady state within 3h, but that RecA bundles continue to form in response to constantly accumulating DNA damage.

Both recA and recA-mGFP alleles were under control of the same native promoter, so a reasonable expectation is that unlabelled and labelled RecA are expressed equally, that we assume for simplicity in the absence of additional proteomics data such as quantitative western blots. However, tagging with GFP and linker requires an extra ~240 codons to the mRNA, taking longer to transcribe and translate (38), and has a greater net probability of stochastic errors including ribosome frameshifting (39), premature termination, or introduction of truncating termination codons, that result in non-functional protein (40–43). Together, these potentially contribute to a reduction in transcription and translation efficiency of the larger labelled gene. Thus, in the merodiploid RecA strain there may be a marginally higher quanity of unlabelled compared to labelled RecA in the cell. We assume for simplicity that MMC does not affect the relative expression of the two alleles, but this may also have some potential effect on expression. Despite these potential issues, the correction factor for the stoichiometry of each individual object is expected to be the same across the cell, since both RecA-mGFP and endogenous RecA have been shown to participate without bias in the formation of both storage bodies and filaments (30).

Our measurements confirm that RecA has a very high concentration in the cytosol of live cells. We observe that untreated cultures have around 11,000 molecules of RecA-mGFP per cell, which increases to 14,000 RecA-mGFP molecules in cells treated with MMC. Of the latter, 35 ± 8% resides in filamentous bundles large enough to resolve in millisecond widefield fluorescence images. Applying the merodiploid correction factor of 2, the total copy number is approximately 23,000 ± 400 RecA in untreated cells, increasing to 28,000 ± 600 in treated cells. While the RecA copy number we estimate in untreated cells exceeds the 4,640 ± 1,908 reported previously by Lesterlin et al (20), our more direct estimations are of similar order and correlate with previous work indicating 2,926-10,377 molecules, with 10,377 in EZ rich medium using a ribosome profiling method (44). Approximately 15,000 RecA per cell in rich medium were reported previously, using semi-quantitative immunoblotting (45); the same study found that RecA copy number increased to 100,000 upon MMC treatment. The discrepancies between our study and others may arise from differences in culture media, growth conditions and MMC dosage.

Unlike RecA, we detected only modest quantities of RecB in untreated cells grown in minimal medium: 13.6 ± 0.5 molecules in tracks, and 126 ± 11 molecules in total per cell based on integrated GFP fluorescence corrected for cellular autofluorescence. Several previous reported find RecB is very scarce - on the order of a dozen molecules per cell. A recent study estimated that there are just 4.9 ± 0.3 RecB per cell using a Halotag fusion allele labelled with HTL-TMR, and 4.5 ± 0.4 per cell using magnetic activiated cell sorting of the same RecB-sfGFP strain we use here, albeit in M9 medium and restricted to nascent cells for which the average copy number is halved (21). An earlier mass spectrometry study used intensity based absolute quantification to estimate 9-20 RecB molecules per cell across different stages of growth in M9 minimal media (46). Ribosome profiling estimated the RecB copy number to be 33-93 molecules per cell in different growth media (44). However, these techniques are either ex vivo or necessitate significantly perturbed intracellular crowding that could result in potentially non-physiological molecular assemblies.

Comparing RecB in tracks in our present study with the number of RecB reported previously, we find a similar if slightly higher estimate, possibly because our approach is based on unsynchronized cultures, as opposed to selecting nascent cells, using fluorescent fusions with a high labelling efficiency. However, our measurement of RecB copy number exceeds previous estimates. The large remainder in integrated fluorescence may represent two possible contributions. The first is from RecB that diffuses faster that our Slimfield microscopy can track - we estimate this limit is approximately 3.5 $\mu m^2$/s, though it is conceivable that free monomeric RecB-sfGFP could exceed this, given that unfused GFP reaches ~8 $\mu m^2$/s in E. coli cytosol (47, 48).The second is not a contribution from RecB at all, but instead a putative artifact of the fusion: an increase in net autofluorescence relative to the parental strain when the real RecB are labelled. However, in our case a threefold increase in autofluorescence would be required to account for our fluorescence measurements, and such a drastic increase in autofluorescence lacks precedent. For example, even stressing cells with MMC sufficiently to induce widespread RecA filamentation only results in an increase in autofluorescence of ~20%. Furthermore, the measured rate of photobleaching of the diffuse RecB-sfGFP signal matches that of RecB-sfGFP trajectories and is roughly half the rate of the autofluorescent parental cells (Figure S4, Table S2). The implication is that untracked RecB-sfGFP is the major contributor to mean cellular fluorescence, which is then a more accurate reflection of total copy of RecB than simply the number of molecules in tracks.

Although the RecB pool is relatively unaffected on MMC treatment, we found that the RecB copy number of RecB in tracks decreased from 13.6 ± 0.5 to 9.5 ± 0.3 molecules per cell. Not only did the average stoichiometry of trajectories decrease by 8% ± 4%, but the number of tracks per cell also dropped due to a sharp increase in the proportion of cells in which RecB assemblies were absent, from 6% to 21%. This reduction in RecB was at odds with our expectation that MMC would eventually lead to increases in DSBs and presumably greater demand for RecB-mediated DSB processing (4). Rather than initiating cellular upregulation of RecB, MMC treatment acts to deplete localized assemblies. Given that DSBs occur in an overwhelming majority of MMC-treated cells as indicated by the induction of RecA filaments, the fate of RecB assemblies cannot simply reflect the presence or absence of DSBs. The increase in the fraction of cells lacking RecB foci follows the expectation under random, independent survival of assemblies (Figure 1J, MMC+ consistent with Poisson distribution with same mean; Pearson $\chi^2$ test, n=234, p=0.012). This result is consistent with a situation where preexisting RecB (hetero)complexes at foci have a higher affinity for dsDNA ends than RecB in the cytoplasmic pool, provided such complexes have a higher rate of disassembly. Such instability might result from successfully bridged pairs of RecA filaments, but we did not detect a correlated loss of pairs of RecB foci. Notably, the number of tracked foci detected per cell is approximately 2 for both RecA and RecB. Could the average of 2 be related to the number of replication sites, or perhaps just reflect a small average number of severe DNA damage sites per cell? Future colocalisation studies of RecA and RecB assemblies may offer more direct insight into the functional interaction and turnover of these repair proteins.

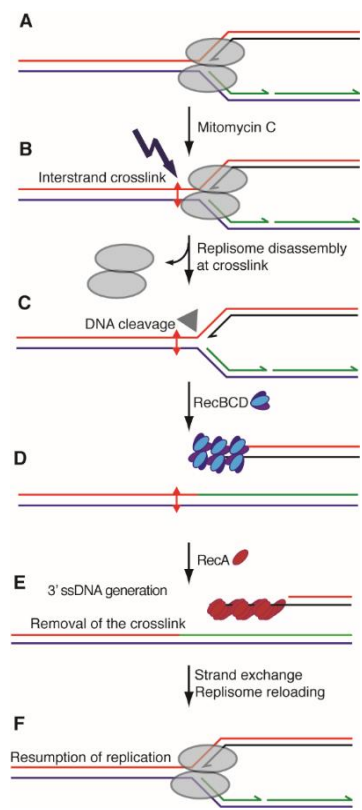

**Figure 5.** A model of DNA damage effected by MMC and subsequent repair at the replication site by RecA and RecBCD. A) Intact replication fork; B) Exposure to MMC and induction of interstrand crosslink that acts as a barrier to an approaching replication fork; C) replisome dissociates if unable to overcome barrier; dissociated fork is recognised by cleavage enzymes that can eventually cause DSBs leading to replication fork collapse; D) newly generated DSB is recognised by RecBCD and processed to generate a 3' single strand end; E) RecA loads onto the newly generated ssDNA; RecA is shown as a short stretch for illustrative purposes but may extend for many thousands of monomeric units over several hundreds of nm of ssDNA, and these filaments may be twisted into bundles. F) strand exchange and replisome reloading restores the fork to resume replication.

Independent of MMC, we observed a consistent dimeric stoichiometry subunit for RecB, suggesting that RecBCD occurs as a pair of heterotrimers *in vivo*. Indeed, earlier *in vitro* studies identified the occurrence of (RecBCD)2 complexes, possibly held together by the nuclease domain of the two RecBCD monomers (12). However, the authors concluded that the monomeric form is functional while the dimeric form is nonfunctional (13). Furthermore, crystallisation of the RecBCD complex for structural studies contained two RecBCD-DNA complexes in the asymmetric unit **(49)**. Were the two complexes in these crystals a coincidence or a physiological consequence? From our observations we cannot determine covalent interactions directly between RecB molecules, but can infer from the strong cotracking correlation that the dimeric form of the complex, (RecBCD)2, occurs in live cells, and that *in vitro* observations of such dimers are carried over from their physiological state. Moreover, our findings are consistent with a greater DSB processivity of assemblies containing multiple pairs of RecB than of isolated RecB in the pool. Making the distinction between monomeric pool RecB and monomeric RecBCD, it is therefore possible that oligomeric RecBCD assemblies associate with DNA DSB *in vivo* but function as individual RecBCD monomers - RecB monomers in the pool would potentially act as a reservoir. A mean stoichiometry of ~6 molecules indicates that RecB foci may occur as assemblies with the diffraction-limited spot width of ~250 nm that comprise roughly three pairs of RecBCD heterotrimers (Figure 5). It would be interesting to estimate the stoichiometries of RecC and RecD in future studies to understand their association in processing DSBs in greater detail.

The RecB copy number decreases slightly upon MMC treatment, indicating that there is only a modest, if any, regulatory response to DNA damage. RecB foci increase neither in number nor stoichiometry, which may indicate that MMC does not induce considerable change in RecB expression on the treatment timescale. A previous study found that RecB expression does not change significantly upon exposure to 0.3 µg/ml MMC (50), comparable to 0.5 µg/ml used in our study. MMC is known to induce the SOS response and cell cycle arrest (4). Since RecB is not an SOS-inducible gene, it follows that the RecB expression does not increase after MMC exposure. These observations also indicate that RecBCD is highly processive to the extent that the cell does not need to increase its RecBCD content to manage increased occurence of DSBs. This situation is different for RecA, which to ensure viability must bind to each ssDNA gap, break and processed end, including those generated from processed DSBs. DSBs are processed by RecBCD to produce ss-ends onto which RecA can load. These RecA-ssDNA complexes induce the SOS response, resulting in further increased expression of RecA (51, 52). This effect accounts for our observed increase in RecA protein copy and the large-scale induction of RecA filaments upon MMC exposure. The observed stoichiometries and RecA and RecB copy numbers upon MMC treatment may thus reflect a consequence of SOS induction of RecA while RecB expression remain steady.

A greater increase in RecA copy numbers and foci compared to RecB could also indicate a significant proportion of single strand breaks and single strand gaps at sites of crosslinks. A lack of redundancy of repair pathways for MMC-induced damage indicates that toxicity cannot be fully accounted for by interstrand crosslinks alone (53). MMC damage repair involves many pathways including nucleotide excision repair (NER), which resolve crosslinks into dsDNA breaks. While previous study reported that NER tends to produce ssDNA nicks (54), we do not know if this applies strictly for MMC-induced NER, as our present work does not pertain to genes that process ss-gaps. Future analysis of proteins that process ssDNA breaks will shed light on the relative occurrence of the two types of breaks by MMC.

We do not know the effect of recB deletion and MMC treatment on RecA dynamics. To avoid RecA interference in 'normal' ssDNA processes such as replication, the cell maintains strict control over filament nucleation, based on RecA concentration and that of its cofactors. It is therefore likely that the observed filamentation upon MMC treatment is dependent on RecBCD, indirectly pointing towards increased occurrences of DSBs in MMC treated cells. Alternatively, if RecA nucleation is independent of RecBCD, one would expect little change in RecA dynamics upon recB deletion. Further analysis of RecA stoichiometry and copy number in a strain devoid of RecBCD activity - and with a controlled RecFOR pathway (17) - could differentiate between these models.

We conclude that RecA occurs as assemblies located near poles of wild type cells. Upon treatment with MMC, RecA assembles into long filamentous bundles on newly generated ssDNA. These long filamentous assemblies may facilitate homology search for homologous base-pairing with an intact duplex. Generation of ssDNA is known to occur at a DSB induced by processing of disassembled forks upon recognition by RecBCD. We observed RecB as a set of three associated dimers at two locations in the cell, providing further evidence that RecBCD predominantly occurs as pairs of heterotrimers inside the cell at either end of DSBs. RecB is not upregulated upon MMC exposure, consistent with it not being a part of the SOS regulon, but instead has a reduced ability to form these periodic assemblies potentially associated with further DSB repair.

## 4. Materials and Methods

### 4.1. Culture and MMC protocol

*E. coli* were grown overnight to mid-log phase in 56-salts minimal media at 30°C, concentrated to ~$10^8$ cells/ml (OD600 ~ 0.3). Aliquots were adjusted to either nil (MMC-) or 0.5 µg/ml MMC (MMC+) and incubated at 30°C for a further 3h. Cells were harvested for microscopy on 1% w/v agarose pads suffused with the same liquid media and imaged within 1h.

### 4.2. Slimfield

A custom-built Slimfield microscope was used for dual-colour, single-molecule-sensitive imaging with a bespoke GFP/mCherry emission channel splitter as described previously (11, 23). The setup included a high-magnification objective (NA 1.49 Apo TIRF 100× oil immersion, Nikon) and the detector was a Prime95B sCMOS camera (Photometrics) operating in 12-bit gain at 200 Hz and 2.4 ms exposure/frame, for a total magnification of 53 nm/pixel. The samples were illuminated in camera-triggered alternating frames by collimated 488 nm or 561 nm wavelength continuous wave OPSL lasers (Coherent, Obis LS) in Gaussian TEM00 mode at a power density of 5 kW/cm2. The number of frames per acquisition was 2,000 for RecA and 300 for RecB strains.

### 4.3. Tracking analysis

Particle tracking employed image processing by custom ADEMscode software in MATLAB (Mathworks) (23, 55). This pipeline determined tracks and mean square displacements from which diffusion coefficients were interpolated. The intensity of each track was estimated by integrating the local pixel values with local sliding window background subtraction. To avoid undercounting bias due to photobleaching, only tracks in the first 10 frames were considered for stoichiometry estimates. The intensities of the constituent foci were extrapolated back to the timepoint of initial laser exposure and divided by the characteristic signal associated with one fluorescent protein under a fixed excitation-detection protocol. After photobleaching sufficiently to show isolated molecules, the characteristic integrated intensity of a single GFP molecule was estimated from the asymptotic distribution of integrated intensity of spots at later timepoints. This direct approach was less feasible for RecA due to its high copy number mandating a prohibitively large number of frames when photobleaching to a strict single-molecule level. More rigorous estimates of the signal per GFP in each dataset were determined from the monomeric intervals in total integrated intensity due to stepwise photobleaching, as identified by a Chung-Kennedy edge-preserving filter (15 ms window, 50% weighting, Figure S1) (56). This integrated intensity is characteristic for each fluorescent protein under fixed imaging conditions, although mGFP and sfGFP were found to be indistinguishable in this respect and are hereafter referred to collectively as GFP. To ensure consistent counts per single-molecule probe, analysis was restricted to the uniformly illuminated area lying within half of the $1/e^2$ beamwaist of the excitation laser. The integrated intensity of GFP in vivo was determined within 14% and 9% respective errors in RecA and RecB (88 ± 18 and 177 ± 16 pixel grey values per GFP for the respective gain modes). The combined equivalent is 88 ± 7 photoelectrons per GFP per frame, which is precise enough to unequivocally identify groups or steps of up to 12 GFP molecules.

Uncorrected cell copy numbers were determined using the CoPro software module with the characteristic GFP intensity and the dark pixel bias as input (23). From this we subtracted the autofluorescence contribution estimated from the parental strain (adjusted by the ratio of mean cell areas).

Photobleaching rates were estimated by fitting the decrease in background-subtracted copy number or mean track stoichiometry over the exposure time using MATLAB cftool. The fit consisted of a monoexponential decay to the first 10 frames with variable initial intensity and decay constant, but with a baseline fixed to the average intensity after 50 frames. Fits were then refined to include only data within the initial $1/e$ decay time (Table S2). RecA-mGFP and RecB-sfGFP photobleach decay times were consistently dissimilar at 13 ± 2 and 6 ± 1 frames respectively; sfGFP is typically several-fold less photostable than comparable enhanced GFPs under high intensity illumination (57).

Author Contributions:

A.P-D, A.S, and M.L. designed the research; A.S. cultured and treated cells; A.P-D performed microscopy and data analysis, visualization and curation, J.S. and L.F. wrote and validated segmentation software; A.P-D and A.S drafted the paper; A.P-D, A.S, J.S. and M.L. edited the paper; M.L. supervised and administered the project. This research was funded by BBSRC, grant numbers BB/P000746/1 and BB/N006453/1, and EPSRC, grant number EP/T002166/1, to M.L.

Data Availability Statement: All raw and analysed data is available on reasonable request from the authors; the MATLAB tracking analysis code can be found at https://github.com/alex-payne-dwyer/single-molecule-tools-alpd. The U-Net image segmentation architecture originated from code obtained from the NEUBIAS Academy workshop (http://eubias.org/NEUBIAS/training-schools/neubias-academy-home/).

Acknowledgments: The authors thank Dr. Christian Lesterlin for the gift of the RecA-mGFP strain and Prof. Meriem El Karoui for the RecB-sfGFP strain.

Conflicts of Interest: The authors declare no conflict of interest. The funders had no role in the design of the study; in the collection, analyses, or interpretation of data; in the writing of the manuscript, or in the decision to publish the results.

**Supplementary Information**

**RecA and RecB: probing complexes of DNA repair proteins with mitomycin C in live Escherichia coli with single-molecule sensitivity**

Alex L. Payne-Dwyer[1,2†], Aisha H. Syeda[1,2†], Jack W. Shepherd[1,2], Lewis Frame[3] and Mark. C. Leake[1,2*]

[1] Department of Physics, University of York, York, UK, YO10 5DD;

[2] Department of Biology, University of York, York, UK, YO10 5DD;

[3] School of Natural Sciences, University of York, York, UK, YO10 5DD.

[*]Correspondence: mark.leake@york.ac.uk;

[†]These authors contributed equally.

|  | RecA-mGFP (~1/2 total RecA) |  | RecB-sfGFP |  |
|---|---|---|---|---|
| MMC treatment | - | + | - | + |
| Number of cells | 190 | 67 | 249 | 307 |
| of which contain tracks | 170 | 60 | 234 | 242 |
| Projected cell area ($\mu m^2$) | 1.8 ± 0.2 | 2.1 ± 0.2 | 1.5 ± 0.1 | 1.6 ± 0.1 |
| Number of tracks | 316 | 125 | 514 | 478 |
| Number of tracks per cell | 1.66 ± 0.06 | 1.86 ± 0.16 | 2.06 ± 0.09 | 1.56 ± 0.06 |
| Number of tracks per cell (with at least one track) | 1.86 ± 0.07 | 2.08 ± 0.17 | 2.19 ± 0.10 | 1.98 ± 0.07 |
| Raw copy, total per cell (molecules) | 11600 ± 200 | 14200 ± 300 | 185 ± 7 | 174 ± 6 |
| Autofluorescence, total per cell (equivalent GFP molecules) | 71 ± 9 | 97 ± 17 | 59 ± 9 | 73 ± 13 |
| Copy, total per cell (molecules) | 11500 ± 200 | 14100 ± 300 | 126 ± 11 | 101 ± 14 |
| Copy number in all tracks per cell (molecules, all cells) | 510 ± 30 | 1080 ± 60 | 13.6 ± 0.5 | 9.5 ± 0.3 |
| Copy numbers in all tracks per cell (molecules, cells with tracks) | 570 ± 30 | 1210 ± 70 | 14.4 ± 0.5 | 12.1 ± 0.3 |
| Track stoichiometry (molecules) | 310 ± 8 | 580 ± 30 | 6.59 ± 0.14 | 6.10 ± 0.19 |
| Raw pool stoichiometry (molecules) | 33.8 ± 4.3 | 33.6 ± 3.7 | 0.69 ± 0.06 | 0.62 ± 0.09 |
| Autofluorescent pool stoichiometry (molecules) | 0.23 ± 0.03 | 0.27 ± 0.04 | 0.23 ± 0.03 | 0.27 ± 0.04 |
| Pool stoichiometry (molecules) | 33.6 ± 4.3 | 33.3 ± 3.7 | 0.46 ± 0.07 | 0.35 ± 0.10 |
| Diffusion coefficient ($\mu m^2$/s) | 0.17 ± 0.02 | 0.07 ± 0.01 | 0.82 ± 0.03 | 0.79 ± 0.03 |
| Track duration (ms) | 39 ± 2 | 89 ± 9 | 31 ± 2 | 32 ± 2 |

**Table S1.** Average properties of molecular tracks for each strain and condition. Values shown as mean ± SEM. Amounts of RecA shown represent only direct measurements of RecA-mGFP and do not account for the similar expression of unlabelled RecA.

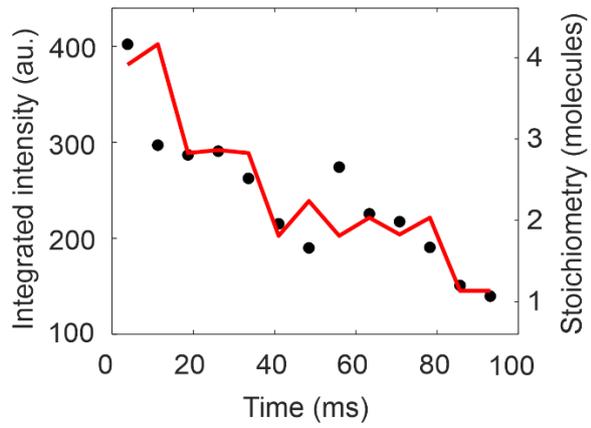

**Figure S1.** Stepwise photobleaching to determine characteristic number of counts per GFP per frame. A single track of RecA-mGFP is shown with integrated intensity in subsequent frames (circles) and the edge-preserving Chung-Kennedy filtered trace, window width 3 frames (line).

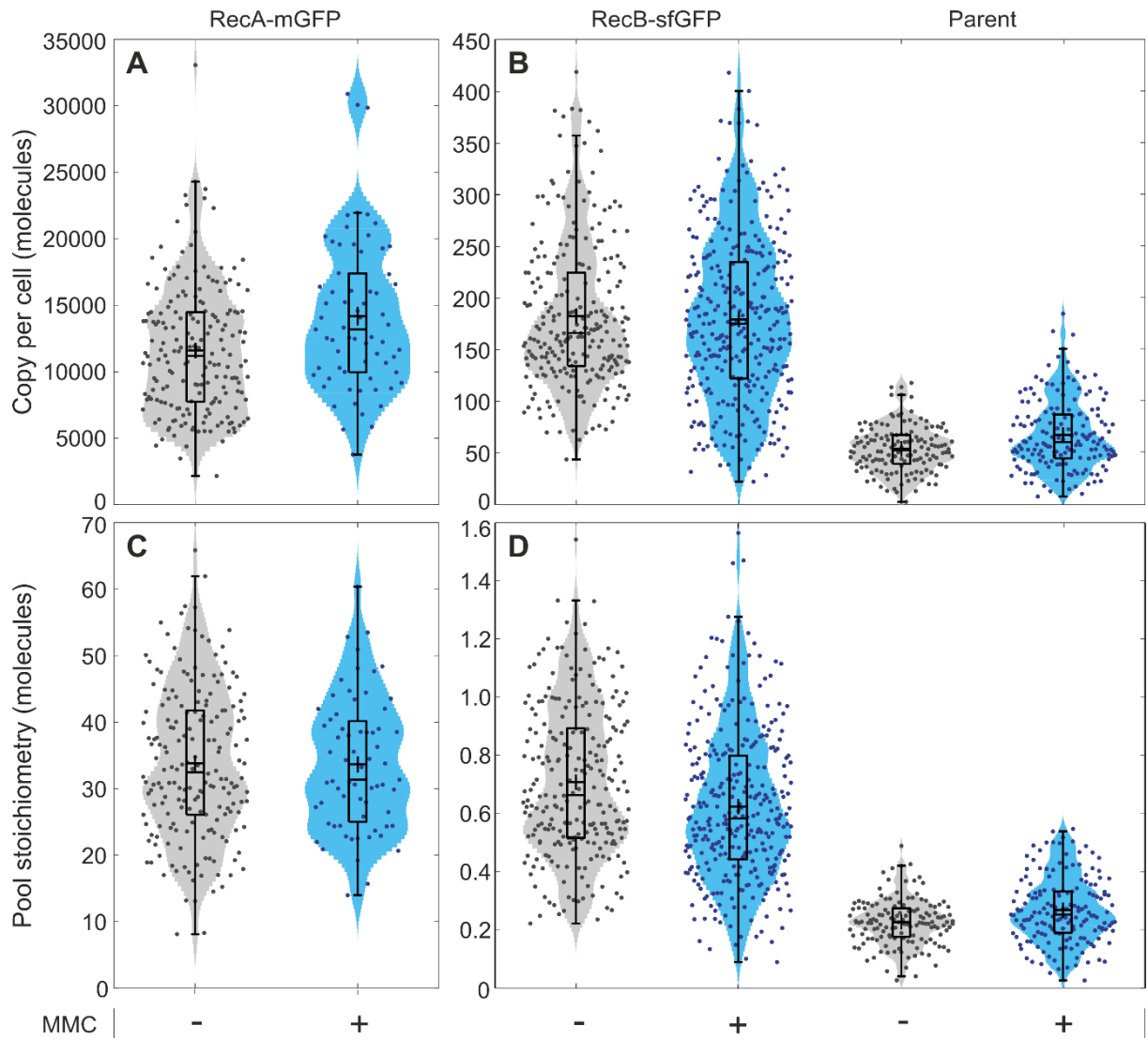

**Figure S2.** Population-level distributions of copy numbers of molecules (GFP equivalents) per cell of A) labelled RecA protein and B) RecB protein *vs.* the unlabelled MG1655 parent strain as determined from total fluorescent intensity. Individual cell data are shown overlaid with the associated 'violin' kernel density distributions and boxplots with interquartile range (box), median (horizontal solid line) and mean with standard error (black cross). By determining the mean concentration of protein outside tracked foci, we find distributions of the upper possible limit of untracked stoichiometry of C) labelled RecA and D) RecB in the intracellular pool, vs. that of the unlabelled parent strain. Both copy and pool stoichiometry metrics are shown for cultures with (blue) and without (grey) MMC treatment. Numbers of cells and statistics as described per condition in Table A1.

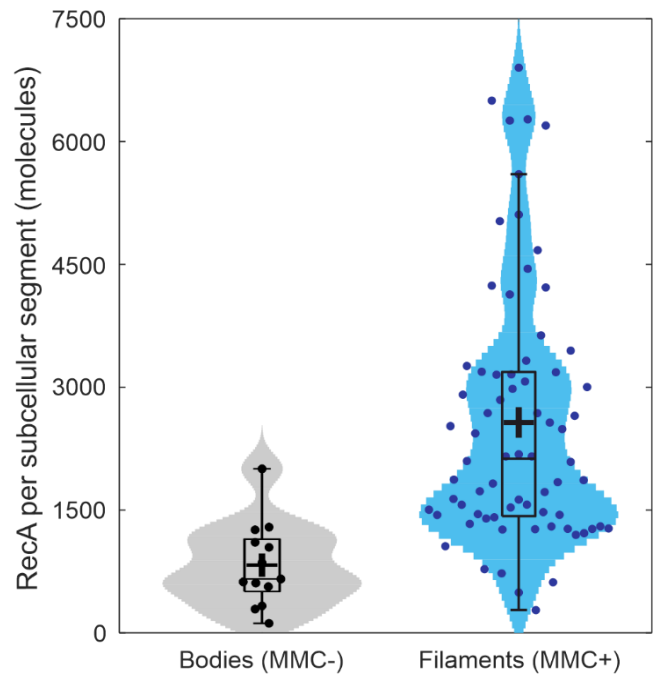

**Figure S3.** Distributions of copy numbers of RecA-mGFP molecules per assembly as determined by local intensity thresholded objects A) in untreated cells, identified with RecA storage bodies, and B) in MMC treated cells, identified with active RecA bundles.

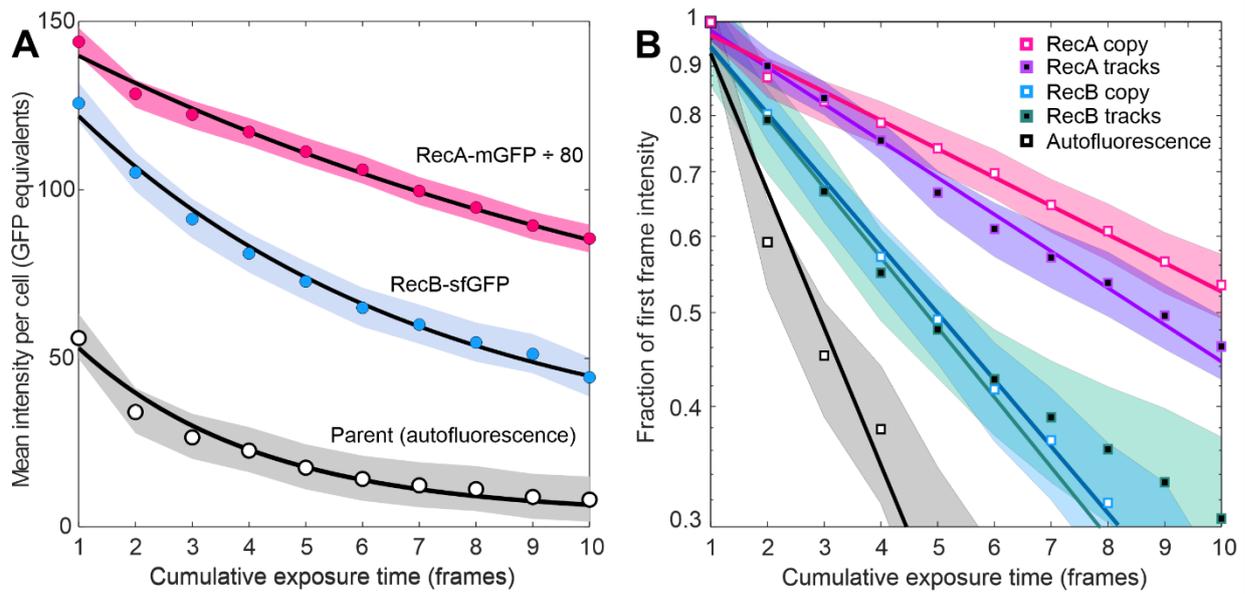

**Figure S4.** A) Photobleaching traces expressed as apparent molecular copy of RecA-mGFP (reduced by a factor of 80 for clarity), RecB-sfGFP and the autofluorescent parent strain over initial exposure timepoints in the absence of MMC. Not only are these distinguishable in magnitude, but they photobleach at different initial rates. Shaded areas are 95% confidence intervals on data at each timepoint. Solid lines shown are single exponential fits to first 10 points unconstrained in height and slope, but with fixed baselines determined after 50 frames. B) Normalised photobleaching traces from copy number, or from mean trajectory brightness over time. The same fits are shown on a logarithmic scale to compare rates more readily. The decay of the excess RecB signal strongly resembles that of RecB confined to trajectories; this indicates correctly labelled RecB-sfGFP is the dominant contribution to the apparent RecB copy rather than increased autofluorescence.

**Table S2**. Characteristic photobleaching decay times (mean ± SEM) as determined from refined monoexponential fits in Figure S4.

| Signal component | Photobleach decay time (frames) |
| --- | --- |
| Autofluorescence | 3.1 ± 0.8 |
| RecA-mGFP | 13.9 ± 1.7 |
| RecA-mGFP in trajectories | 12.3 ± 1.8 |
| RecB-sfGFP | 6.2 ± 1.1 |
| RecB-sfGFP in trajectories | 6.0 ± 0.9 |

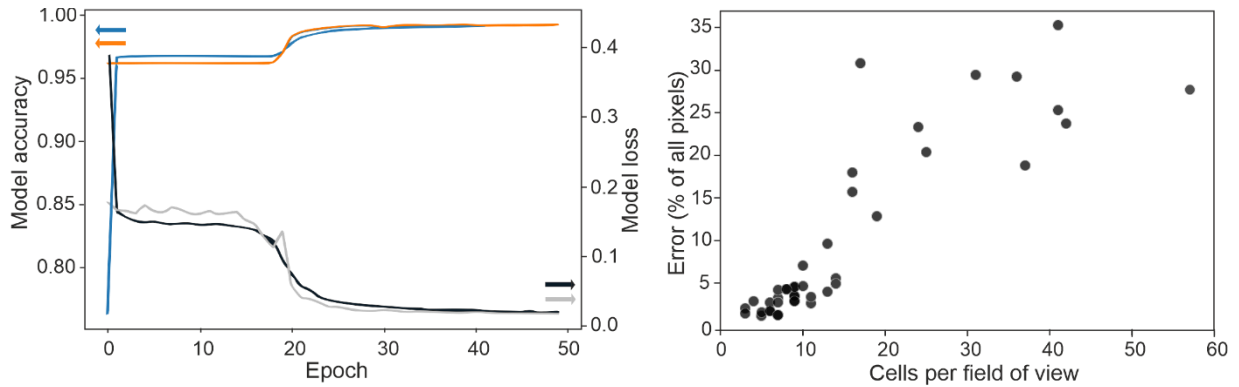

**Figure S5:** A) The accuracy and loss of the segmentation model plotted against iterative epoch for the training images (blue, black respectively) and validation images (orange, grey respectively) using the U-Net with optimal parameters; B) rate of classification error for pixels in image against number of cells per field of view for all experimental datasets.

**Supplementary Methods**

*Segmentation and image post-processing*

Our in-house manual segmentation protocol in MATLAB was used to identify individual cells from widefield images. To improve throughput and reproducibility, we also developed a complementary AI-driven method to segment cell-containing image pixels from background pixels. This method consisted of a standard U-Net convolutional neural network with four layers (58), followed by object labelling using CellProfiler (59). The U-Net was trained with a binary cross entropy loss function (60) and an Adam optimizer (61) for 50 epochs with 10% dropout for the top two layers, 20% dropout for the bottom two layers, and a batch size of 6. We used 248 training images with a 9:1 training:validation split and reserved 76 images for testing. Images were prepared by registration of the brightfield and fluorescence micrographs and were constrained to be 256x256 pixels in size, roughly the size of our epifluorescence excitation spot. The best performance was obtained with a learning rate of $2.5\times10^{-4}$ and a patience value of 7 epochs to avoid overfitting (62), giving an accuracy of 0.995 for all image pixels on our validation and testing sets (Supplementary Figures S5A and S5B). For real-life tests we compared our U-Net segmentation of fields of view with varying densities of cells with MATLAB hand-segmentations of the same image. We find that in cases of low cell number density, the segmentation classifies all pixels with good (>95%) accuracy (Figure S5C). As cell number density increases, the pixel classification accuracy decreases (towards ~70%) due to ambiguity at cell boundaries and contacts. This trend is to be expected since the brightfield training sets include primarily planktonic E. coli at low contrast. Indeed, our choice of ground truth is not itself immune to this ambiguity at cell contacts, which at the cost of convenience, would be better specified with a membrane-localised fluorescent control. Nonetheless, this proof-of-concept implementation is more effective than pretrained or unsupervised methods for the Slimfield modality to date. To achieve a fully automated pipeline with true object labelling, we will incorporate additional postprocessing steps such as denoising, background subtraction, and removal of spurious cell boundaries (63).